\documentclass[12pt]{iopart}

\begin{document}

\title{$\Lambda$CDM epoch reconstruction from F(R,G) and modified Gauss-Bonnet gravities}

\author{E Elizalde$^1$,  R Myrzakulov$^2$, V V Obukhov$^3$ and D S\'aez-G\'omez$^1$}

\address{$^1$Consejo Superior de Investigaciones Cient\'{\i}ficas ICE/CSIC-IEEC, Campus UAB, Facultat de Ci\`encies, Torre C5-Parell-2a pl, E-08193 Bellaterra (Barcelona) Spain}
\address{$^2$Dept. Gen. Theor. Phys., Eurasian National University,  Astana, 010008, Kazakhstan}
\address{$^3$Tomsk State Pedagogical University, Tomsk, Russia}

\ead{elizalde@ieec.uab.es, elizalde@math.mit.edu}

\begin{abstract}
Dark energy cosmology is considered in a modified Gauss-Bonnet model of gravity with and without a scalar field. It is shown that these generalizations of General Relativity endow it with a very rich cosmological structure: it may naturally lead to an effective cosmological constant, quintessence or phantom cosmic acceleration, with the possibility to describe the transition from a decelerating to an accelerating phase explicitly. It is demonstrated here that these modified GB and scalar-GB theories are perfectly viable as cosmological models. They can describe the $\Lambda$CDM cosmological era without any need for a cosmological constant. Specific properties of these theories of gravity in different particular cases, such as the de Sitter one, are studied.
\end{abstract}


\maketitle

\section{Introduction}

Recent observational data indicate that our universe is accelerating. This acceleration is explained in terms of the so-called dark energy (DE), which may be explained in modified gravity models (for a general review see Ref.\cite{N1}). DE could also result from a cosmological constant, from an ideal fluid with a different form of equation of state and negative pressure, a scalar field with quintessence-like or phantom-like behavior (see \cite{Eli1} and references therein), etc. The choice of possibilities reflects the indisputable fact that the true nature and origin of DE has not been convincingly explained yet. It is not even clear what type of DE is more seemingly to explain the current epoch of the universe. Observational data point towards some kind of DE with an equation of state (EoS) parameter which is very close to -1, maybe even less than -1 (the so-called phantom case). A quite appealing possibility is the already mentioned modification of General Relativity (GR). Modifications of the Hilbert-Einstein action by introducing different functions of the Ricci scalar have been systematically explored, the so-called $F(R)$ gravity models, which reconstruction has been developed in Refs.~\cite{N7}-\cite{DSG}. 
As is known, $F(R)$  gravity can be written in terms of a scalar field—quintessence or phantom like—by redefining the function $F(R)$  with the use of a scalar field, and then performing a conformal transformation. It has been shown that, in general, for any given $F(R)$ the corresponding scalar-tensor theory can, in principle, be obtained, although
the solution is going to be very different from one case to another. Also, attention has been paid to the reconstruction of $F(R)$ gravity from a given scalar-tensor theory. It is known, too, that the phantom case in scalar-tensor theory does not exist, in general, when starting from $F(R)$ gravity. In fact, the conformal transformation becomes complex
when the phantom barrier is crossed, and therefore the resulting $F(R)$ function becomes complex. These situations where addressed in \cite{DSG} in detail, where to avoid this hindrance, a dark fluid was used in order to produce the phantom behavior in such a way that the $F(R)$ function reconstructed from the scalar-tensor theory continues to be real.

On the other hand, it has also been suggested in the literature (see Refs.~\cite{N2}, \cite{N9}) to consider modified Gauss-Bonnet gravity, that is, a function of the GB invariant. Different cosmological properties of modified gravity models of this kind have been studied in Refs.~\cite{N2}-\cite{Bamba}. Both possibilities have, in principle, the capability to explain the accelerated expansion of the Universe and even the primordial inflationary phase (see \cite{N7} and \cite{DSG}), with no need to introduce a new form of energy. In this paper we will study some specific modified Gauss-Bonnet theories and, by using a technique developed in Ref.~\cite{N4}, the corresponding cosmological theory will be reconstructed for several cosmological solutions.

We will here work with the spatially-flat FRW universe metric 
\begin {equation}
ds^{2}=-dt^{2}+a(t)^{2}\sum^{3}_{i=1}(dx^{i})^{2},
\label{FRW}
\end{equation}
where $a(t)$ is the scale factor at cosmological time $t$. In GR, the corresponding system of equations are the usual Friedmann equations, namely
\begin{equation}
\frac{3}{k^2}H^2=\rho, \quad \frac{1}{k^2}(2\dot{H}+3H^2)=-p, \quad \dot{\rho}=-3H(\rho+p).
\label{I1}
\end{equation}
The last of them is the continuity equation for a perfect fluid. As usually, the Hubble rate $H$ is defined by $H\equiv\dot{a}/a$. In (\ref{I1}), $\rho$ and $p$ are the matter energy-density and pressure. The Gauss-Bonnet invariant is given by
\begin{equation}
 G=R^2-4R_{\mu\nu}R^{\mu\nu}+R_{\mu\nu\lambda\sigma}R^{\mu\nu\lambda\sigma}
\label{I2}
\end{equation}
The cosmological models coming from the different versions of modified GB gravity here considered will be carefully investigated with the help of several particular examples
where calculations can be carried out explicitly.

The paper is organized as follows. In Sect.~II, the general technique which will be used to reconstruct a gravity theory from a given cosmological evolution setup will be explained in detail. Several specific examples will be worked through, for the simple case where one considers a Hilbert-Einstein action plus a function on the GB invariant. In Sect.~III, a general model whose action depends on a function of the Ricci scalar and of the GB invariant will be studied. With the help of several explicit examples we will show how the reconstruction of the cosmological solutions is carried out. In Sect.~IV the more involved case of a scalar-tensor theory where the GB invariant is coupled to the scalar field will be considered. It will be shown there that any cosmological solution can be reproduced by an specific scalar potential and a convenient coupling term. Finally, in the conclusions we will provide a summary of the main results obtained in the paper.

\section{The $[R+f(G)]$-model}

Consider the following action, which describes General Relativity plus a function of the Gauss-Bonnet term (see Refs.~\cite{N2} and \cite{N9}),
\begin {equation}
S=\int d^{4}x\sqrt{-g}\left[\frac{1}{2\kappa^{2}}R+f(G)+L_{m}\right]\ ,
\label{1.1}
\end{equation}
where $\kappa^2=8\pi G_N$, $G_N$ being the Newton constant. By varying the action over $g_{\mu\nu}$, the following field equations are obtained
\begin{eqnarray}
\fl 0=\frac{1}{2k^2}(-R^{\mu\nu}+\frac{1}{2}g^{\mu\nu}R)+T^{\mu\nu}+\frac{1}{2}g^{\mu\nu}f(G)-2f_{G}RR^{\mu\nu}+4f_{G}R^{\mu}_{\rho}R^{\nu\rho}-2f_{G}R^{\mu\rho\sigma\tau}R^{\nu}_{\rho\sigma\tau}\nonumber\\
-4f_{G}R^{\mu\rho\sigma\nu}R_{\rho\sigma}+2(\nabla^{\mu}\nabla^{\nu}f_{G})R
-2g^{\mu\nu}(\nabla^{2}f_{G})R-4(\nabla_{\rho}\nabla^{\mu}f_{G})R^{\nu\rho}-4(\nabla_{\rho}\nabla^{\nu}f_{G})R^{\mu\rho} \nonumber\\ +4(\nabla^{2}f_{G})R^{\mu\nu}+4g^{\mu\nu}(\nabla_{\rho}\nabla_{\sigma}f_{G})R^{\rho\sigma}-4(\nabla_{\rho}\nabla_{\sigma}f_{G})R^{\mu\rho\nu\sigma}.
\label{1.2}
\end{eqnarray}
For the metric (\ref{FRW}), these equations give the first FRW equation which has the form
\begin {equation}
0=-\frac{3}{k^{2}}H^{2}+G f_{G}-f-24\dot{G}H^{3}f_{GG}+\rho_{m}.
\label{1.3}
\end{equation}
The Hubble rate $H$ is here defined by $H=\dot{a}/a$, while 
$G$ and  $R$ are given by
\begin {equation}
G=24(\dot{H}H^{2}+H^{4}),\quad R=6(\dot{H}+2H^{2}).
\label{1.4}
\end{equation}
The matter energy density $\rho_{m}$ satisfies the standard continuity equation
\begin {equation}
\dot{\rho_{m}}+3H(1+w)\rho_{m}=0.
\label{1.5}
\end{equation}
Let us  now rewrite Eq.~(\ref{1.3}) by using a new variable
$N=\ln\frac{a}{a_{0}}=-\ln(1+z)$, that is, the number of e-foldings, instead of
the cosmological time $t$, where  $z$ is the redshift (this method has been implemented in Ref.~\cite{N4} for $f(R)$ gravity). The following expressions are then easily obtained
\begin{eqnarray}
a=a_{0}e^{N},\quad H=\dot{N}=\frac{dN}{dt}, \quad
\frac{d}{dt}=H\frac{d}{dN},  \nonumber\\
\frac{d^2}{dt^2}=H^2\frac{d^2}{dN^2}+HH^\prime\frac{d}{dN},\quad
H^{\prime}=\frac{dH}{dN}.
\label{1.6}
\end{eqnarray}
Eq.~(\ref{1.3}) can thus be expressed as follows
\begin{eqnarray}
0=-\frac{3}{k^{2}}H^2+24H^{3}(H'+H)f_{G}-f\nonumber\\
-576H^{6}\left(HH''+3H'^{2}+4HH'\right)f_{GG}+\rho_{m}\ ,
\label{1.7}
\end{eqnarray}
where $G$ and $R$ are now
\[
G=24(H^{3}H'+H^{4}),
\quad\dot{G}=24(H^{4}H''+3H^{3}H'^{2}+4H^{4}H'), 
\]
\begin{equation}
 R=6(HH'+2H^{2}).
\label{1.8}
\end{equation}
By introducing a new function $g$ as $g=H^2$, we have
\begin {equation}
 H=\sqrt{g},\quad  H'=\frac{1}{2}g^{-1/2}g',\quad
H''=-\frac{1}{4}g^{-3/2}g'^{2}+\frac{1}{2}g^{-1/2}g''.\end{equation}
Hence,  Eq.~(\ref{1.7})  takes the form
\begin{equation}
0=-\frac{3}{k^{2}}g+12g(g'+2g)f_{G}-f-24^{2}g\left[\frac{1}{2}g^{2}g''
+\frac{1}{2}gg'^{2}+2g^{2}g'\right]f_{GG}+\rho_{m}\ ,
\label{1.9}
\end{equation}
where we have used the expressions
\begin{eqnarray} G=12gg'+24g^{2},\quad
\dot{G}=12g^{-1/2}[g^2g''+gg'^2+4g^{2}g']\ , \nonumber\\ 
\quad R=3g'+12g.
\label{1.10}
\end{eqnarray}

Finally, we can write the FRW equations in a slightly different form \cite{Cog1}
\begin{equation}
\frac{3}{k^2}H^2=\rho_{eff}\ , \quad \frac{1}{k^2}(2\dot{H}+3H^2)=-p_{eff},
\label{1.11}
\end{equation}
where the effective energy and pressure densities are 
\begin{equation}
p_{eff}=w_{eff}\rho_{eff}, \quad  \dot{\rho}_{eff}=-3H(\rho_{eff}+p_{eff}),
\label{1.12}
\end{equation}
with
\begin{equation}
\rho_{eff}=\rho_{G}+\rho_{m}, \quad p_{eff}=p_{G}+p_{m}.
\label{1.13}
\end{equation}
Here
\begin{eqnarray}
\rho_{G}=Gf_{G}-f-24H^3\dot{G}f_{GG}, \nonumber\\
p_{G}=-\rho_{G}+8H^2\dot{G}^2f_{GGG}\nonumber\\
-192f_{GG}(4H^6\dot{H}-8H^3\dot{H}\ddot{H}-6H^2\dot{H}^3-H^4\tdot{H}-3H^5\ddot{H}-18H^4\dot{H}^2).
\label{1.14}
\end{eqnarray}
As it was pointed out in Ref.~\cite{Cog1}, for de Sitter space $p_{G}=w_{G}\rho_{G}=-\rho_{G}$, that is $w_{G}=-1$.

\subsection{Example 1}

As a first example we consider the $\Lambda$CDM-model. As will be shown below, for the gravity theory described by the action (\ref{1.1}), it is possible to reconstruct such evolution with no need of a cosmological constant term.
For the $\Lambda$CDM case, we have
\begin{equation}g=H^{2}= H^{2}_{0}+\frac{k^{2}\rho_0}{3}a^{-3}=H^{2}_{0}+\frac{k^{2}\rho_0}{3}a_{0}^{-3}e^{-3N}=
H^{2}_{0}+lz,
\label{1.15}
\end{equation}
where $l=\frac{k^{2}\rho_0a_{0}^{-3}}{3},\quad z=e^{-3N}$. Hence,
we get
\begin{equation}z^{'}=-3z,\quad  g'=-3lz,\quad
g''=9lz.
\label{1.16}
\end{equation}
Finally, for $G$ we obtain the expressions
\begin{equation}
G=24H_0^4+12H^2_0lz-12l^2z^2\ , \quad \dot{G}=-3zHW=-3Hz(12H^2_0l-24l^2z),
\label{1.17}
\end{equation}
where $W=12H^2_0l-24l^2z$. We can reverse Eq.~(\ref{1.17}) and the number of e-foldings can be written as a function of the Gauss-Bonnet term. This yields
 \begin{equation}
z=\frac{3H^{2}_{0}\pm\sqrt{81H^{4}_{0}-3G}}{6l}=e^{-3N} \rightarrow N=\frac{1}{3}\ln{\frac{6l}{3H^{2}_{0}\pm\sqrt{81H^{4}_{0}-3G}}}.
\label{1.18}
\end{equation}
Hence, it follows that
\begin{equation}
g=H^{2}=H^{2}_{0}+\frac{3H^{2}_{0}\pm\sqrt{81H^{4}_{0}-3G}}{6}=\frac{9H^{2}_{0}\pm\sqrt{81H^{4}_{0}-3G}}{6}.
\label{1.19}
\end{equation}
Finally, for the function $f(z)$ we get the following equation
\begin{equation}
a_{2}f_{zz}+a_{1}f_z+a_{0}f+b=0,
\label{1.20}
\end{equation}
where
\begin{eqnarray}
a_{0}=-w^2=-24^2l^4z^2+24^2H_0^2l^3z-144H_0^4l^2,\nonumber\\
a_{1}=12(H_0^2+lz)[144l^2z(H_0^2+lz)+(2H_0^2-lz)w],\nonumber\\
a_{2}=72zw(H^2_0+lz)^2, \quad    b=[\rho_{m}-\frac{3}{k^{2}}(H_0^2+lz)]w^2.
\label{1.21}
\end{eqnarray}
The energy density can be expressed as
\begin{equation}
\rho_{m}=-Dz-E,
\label{1.22}
\end{equation}
being
\begin{equation}
D=144HH^4_0l^3[\frac{83}{5k^2}-\frac{16}{9}+\frac{20\delta}{9}],\quad E=144HH^4_0l^2[\frac{4}{9}-\frac{13}{5k^2}-\frac{5\delta}{9}].
\label{1.23}
\end{equation}
Here $\delta=const$. As a consequence, Eq.~(\ref{1.20}) has the following particular solution
\begin{equation}
f(z)=\theta z^2+\vartheta z +H^2_0\delta,
\label{1.24}
\end{equation}
where
\begin{equation}
\theta=\frac{l^2}{H^2_0}[k^{-2}-\frac{2}{9}(\delta+1)], \quad \vartheta=l[\frac{1}{5}k^{-2}+\frac{2}{9}(\delta+1)].
\label{1.25}
\end{equation}
Thus, we have
\begin{equation}
f(G)=\theta [\frac{3H^{2}_{0}\pm\sqrt{81H^{4}_{0}-3G}}{6l}]^2+\vartheta [\frac{3H^{2}_{0}\pm\sqrt{81H^{4}_{0}-3G}}{6l}] +H^2_0\delta.
\label{1.26}
\end{equation}
We observe that this function reproduces exactly the same behavior as the $\Lambda$CDM model in the context of Gauss-Bonnet gravity.

\subsection{Example 2}

Let us now consider a second example
\begin{equation}
g=H^{2}= le^{-3N}=lz.
\label{1.27}
\end{equation}
This case describes a time evolution given by
\begin{equation}
 H(t)=\frac{1}{t-t_0}\ ,
\label{1.27a}
\end{equation}
which is equivalent to the cosmological evolution of a pressureless fluid in  GR, what gives a decelerated expansion. As in the example above, we find that $z=\pm i\frac{\sqrt{3G}}{6l}$. In the present case, we have the following differential equation,
\begin{equation}
z^2f_{zz}-\frac{7}{6}zf_z+\frac{1}{3}f-\frac{c}{3}=0,
\label{1.28}
\end{equation}
where $c=\rho_{m}-\frac{3}{k^{2}}lz$.

Let us assume that $c=0$,  then Eq.~(\ref{1.28}) takes the form
\begin{equation}
z^2f_{zz}-\frac{7}{6}zf_z+\frac{1}{3}f=0.
\label{1.29}
\end{equation}
This equation admits the following exact solution
\begin{equation}
f_1(z)=C_{1}z^{2}+C_{2}z^{\frac{1}{6}}.
\label{1.30}
\end{equation}
As a consequence, for this second example we obtain the following model
\begin{equation}
f_(G)=-C_{1}\frac{G}{12}+C_{2}(\frac{\pm i\sqrt{3}}{6l})^{\frac{1}{6}}G^{\frac{1}{12}}=C_{1}'G+C_{2}'G^{\frac{1}{12}}.
\label{1.31}
\end{equation}
It is now easy to see from this example that the GB terms could actually 
contribute in the matter dominated epoch of the universe evolution, provided
the $f(G)$ function is taken properly, what is certainly an interesting result.

\subsection{Example 3}

As third and last example we will now consider the case when the 
Hubble parameter and the matter energy density behave as
\begin{equation}
H=e^{mN}, \quad \rho_{m}=b_1+\frac{3}{k^2}e^{2mN}+\frac{96(m+1)}{5}b_{2}e^{5mN},
\label{2.3.1}
\end{equation}
where $b_{j}=const$. This case corresponds to the so-called phantom behavior, what means that we have an effective EoS parameter $w_{eff}<-1$, and produces a superaccelerated phase that could end in some kind of future singularity (see Refs.~\cite{Bamba} and \cite{Cap2}). We first write the Hubble parameter as a function of time,
\begin{equation}
H(t)=\frac{H_0}{t_s-t}\ ,
\label{1.32}
\end{equation}
where $t_s$, usually called the Rip time, defines the moment when the Big Rip singularity occurs. We can find the $f(G)$ function that reproduces this model by solving the equation (\ref{1.7}). This yields following solution
 \begin{eqnarray}
F(G)=F(N)=b_{1}+\frac{96(m+1)}{5}b_{2}e^{5mN}=\nonumber\\
=b_{1} +\frac{96(m+1)}{5}b_{2}[\frac{G}{24(m+1)}]^{\frac{5}{4}}.
\label{2.3.2}
\end{eqnarray}

\section{The $F(R,G)$  model}

Let us now consider a more general model for a kind of modified Gauss-Bonnet gravity. 
This can be described by the following action
\begin{equation}
S=\int d^{4}x\sqrt{-g}[\frac{1}{2k^2}F(R,G)+L_{m}].
\label{F(R,G)action}
\end{equation}
Varying over $g_{\mu\nu}$ the gravity field equations are obtained \cite{Cog1},
\begin{eqnarray}
0=T^{\mu\nu}+\frac{1}{2}g^{\mu\nu}F(G)-2F_{G}RR^{\mu\nu}+4F_{G}R^{\mu}_{\rho}R^{\nu\rho}\nonumber\\
-2F_{G}R^{\mu\rho\sigma\tau}R^{\nu}_{\rho\sigma\tau}-4F_{G}R^{\mu\rho\sigma\nu}R_{\rho\sigma}+2(\nabla^{\mu}\nabla^{\nu}F_{G})R
-2g^{\mu\nu}(\nabla^{2}F_{G})R\nonumber\\
-4(\nabla_{\rho}\nabla^{\mu}F_{G})R^{\nu\rho}-4(\nabla_{\rho}\nabla^{\nu}F_{G})R^{\mu\rho}+4(\nabla^{2}F_{G})R^{\mu\nu}+4g^{\mu\nu}(\nabla_{\rho}\nabla_{\sigma}F_{G})R^{\rho\sigma}\nonumber\\
-4(\nabla_{\rho}\nabla_{\sigma}F_{G})R^{\mu\rho\nu\sigma}-F_{G}R^{\mu\nu}+\nabla^{\mu}\nabla^{\nu}F_{R}-g^{\mu\nu}\nabla^{2}F_{R}.
\label{FieldEq}
\end{eqnarray}
In the case of a flat FRW Universe, described by the metric (\ref{FRW}), the first FRW equation yields
\begin {equation}
0=\frac{1}{2}(GF_{G}-F-24H^{3}F_{Gt})+3(\dot{H}+H^{2})F_{R}-3HF_{Rt}+k^2\rho_{m}.
\label{FriedmEq}
\end{equation}
And from here, using the techniques developed in the above section, it is plain that explicit $F(R,G)$ functions can be reconstructed for given cosmological solutions.

\subsection{De Sitter Solutions}

As well known, the de Sitter solution is one of the most important cosmological solutions nowadays, since the current epoch has been observed to have an expansion that behaves approximately as de Sitter. This solution is described by an exponential expansion of the scale factor, which gives a constant Hubble parameter $H(t)=H_0$. By inserting it in the Friedmann equation (\ref{FriedmEq}), one finds that any kind of $F(R,G)$ function can possibly admit de Sitter solutions, with the proviso that the following algebraic equation has positive roots for $H_0$
\begin{equation}
0= \frac{1}{2}(G_0F_G(G_0)-F(G_0,R_0))+3H^2_0F_R(R_0)\ ,
\label{deSitter}
\end{equation}
being $R_0=12H^2_0$ and $G_0=24H^4_0$; we have here neglected the contribution of matter for simplicity. As it was pointed in Refs.~\cite{f(R)deSitter} and \cite{DSG} for the case of modified $F(R)$ gravity, the de Sitter points are critical points for the Friedmann equations, what could explain the current acceleration phase as well as the inflationary epoch. This explanation can be extended to the action (\ref{F(R,G)action}), so that any kind of function $F(R,G)$ with positive real roots for the equation (\ref{deSitter}) could in fact explain the acceleration epochs of the Universe in exactly the same way a cosmological constant does.

\subsection{Phantom dark energy}

Let us now explore the cosmic evolution described by (\ref{2.3.1}) in the context of the action (\ref{F(R,G)action}). This solution  reproduces a phantom behavior, i.e. a superaccelerated expansion that, as pointed by recent observations, our Universe could be in---or either close to cross the phantom barrier. We can now proceed with the reconstruction method, as explicitly shown in the section above, and a $F(R,G)$ function for the Hubble parameter (\ref{2.3.1}) will be obtained. For simplicity, we consider the following subfamily of functions
\begin{equation}
F(R,G)=f_1(G)+f_2(R)\ .
\label{3.1a}
\end{equation}
Correspondingly, the Friedmann equation (\ref{FriedmEq}) can be split 
into two equations, as
\begin{eqnarray}
 0=-24H^3\dot{G}f_{1GG}+ Gf_{1G}-f_1\ , \nonumber\\
0=-3H\dot{R}f_{2RR}+3(\dot{H}+H^2)f_{2R}-\frac{1}{2}f_2 +\kappa^2\rho_m
\label{3.1}
\end{eqnarray}

For the example (\ref{2.3.1}), that is $H=e^{mN}$, the Ricci scalar and the Gauss-Bonnet terms take the following form,
\begin {eqnarray}
G=24(m+1)e^{4mN}=24(m+1)H^4,\nonumber\\ 
R=6(m+2)e^{2mN}=6(m+2)H^2\ .
\label{3.2}
\end{eqnarray}
Hence, the first equation in (\ref{3.1}) can be written in terms of $G$, and this yields
\begin{equation}
 G^2f_{1GG}-\frac{m+1}{4m}Gf_{1G}+\frac{m+1}{4m}f_1=0\ .
\label{3.3}
\end{equation}
This is an Euler equation, easy to solve, and yields
\begin{equation}
f_1(G)=C_1G^{1+\frac{m+1}{4m}}+C_2\ ,
\label{3.4}
\end{equation}
where $C_{1,2}$ are integration constants. In the same way, for the case being considered here, the second equation in (\ref{3.1}), for $R$, takes the form
\begin{equation}
R^2f_{2RR}-\frac{m+1}{2m}Rf_{2R}+\frac{m+2}{2m}f_2-\frac{\kappa^2(m+2)}{m}\rho_m=0\ .
\label{3.5}
\end{equation}
In absence of matter ($\rho_m=0$) this  is also an Euler equation, whose solution yields
\begin{eqnarray}
f_2(R)=k_1R^{\mu_+}+k_2R^{\mu_-}\, \nonumber\\ 
\mbox{where} \quad \mu_{\pm}=\frac{1+\frac{m+1}{2m}\pm\sqrt{1+\frac{(m+1)^2}{4m^2}-\frac{m+3}{m}}}{2}\ ,
\label{3.6}
\end{eqnarray}
and $k_{1,2}$ are integration constants. Then, the complete function F(R,G), given in (\ref{3.1a}), is reconstructed (in absence of matter) yielding the solutions (\ref{3.4}) and (\ref{3.6}). The theory (3.12) belongs to the class of models with positive and negative powers of the curvature introduced in \cite{no_n1}.

Let us now consider the case where matter is included. From the energy conservation equation (\ref{1.5}) we have that, for a perfect fluid with constant EoS, $p_m=w_m\rho_m$,  the solution is given by $\rho_m=\rho_0 e^{-3(1+w_m)N}$.  By inserting the expression for $R$ (\ref{3.2}) into this solution, we get
\begin{equation}
\rho_m=\rho_{m0}=\left(\frac{R}{6(m+2)} \right)^{-\frac{3(1+w_m)}{2m}}\ .
\label{3.7}
\end{equation}
In such case the general solution for $f_2$ is given by
\begin{eqnarray}
f_2(R)=k_1R^{\mu_+}+k_2R^{\mu_-}+kR^A\ , \mbox{where,} \nonumber\\   k=\kappa^2\frac{\rho_{m0}(6(m+1))^{-A}}{A(A-1)-1/2m} \quad \mbox{and} \quad A=-\frac{3(1+w_m)}{2m}\ .
\label{3.8}
\end{eqnarray}
Hence, we see that the solution for the Hubble parameter (\ref{2.3.1}) can be easily recovered in the context of modified Gauss-Bonnet gravity. Nevertheless, it seems clear that, for more complex examples, one may not be able to solve the corresponding equations analytically and it could require numerical analysis.

\section{Reconstruction of scalar-GB gravity}

In this section we consider a four-dimensional action containing the Einstein-Hilbert part, a massless scalar field, and the Gauss-Bonnet term coupled to the scalar field. The corresponding action is 
\begin{equation}
S=\int d^{4}x\sqrt{-g}[\frac{1}{2k^2}R-\frac{1}{2}\partial_{\mu}\phi\partial^{\mu}\phi-V(\phi)-\xi(\phi)G+L_{m}].
\label{4.1}
\end{equation}
An action of this kind has been proposed as a model for dark energy in Ref.~\cite{N10}. It can actually be related with modified GB gravity, as explained in Refs.~\cite{N3} and \cite{Bamba2}. The variation of this action over the metric $g_{\mu\nu}$ gives \cite{N5}
\begin{eqnarray}
0=\frac{1}{k^2}(-R^{\mu\nu}+\frac{1}{2}g^{\mu\nu}R)+T^{\mu\nu}+\frac{1}{2}\partial^{\mu}\phi\partial^{\nu}\phi-\frac{1}{4}g^{\mu\nu}\partial_{\rho}\phi\partial^{\rho}\phi\nonumber\\
+\frac{1}{2}g^{\mu\nu}(-V+\xi G)-2\xi RR^{\mu\nu}-4\xi R^{\mu}_{\rho}R^{\nu\rho}
-2\xi R^{\mu\rho\sigma\tau}R^{\nu}_{\rho\sigma\tau}\nonumber\\
+4\xi R^{\mu\rho\nu\sigma}R_{\rho\sigma}+2(\nabla^{\mu}\nabla^{\nu}\xi)R
-2g^{\mu\nu}(\nabla^{2}\xi)R-4(\nabla_{\rho}\nabla^{\mu}\xi)R^{\nu\rho}
\nonumber\\
-4(\nabla_{\rho}\nabla^{\nu}\xi)R^{\mu\rho}+4(\nabla^{2}\xi)R^{\mu\nu}+4g^{\mu\nu}(\nabla_{\rho}\nabla_{\sigma}\xi)R^{\rho\sigma}+4(\nabla_{\rho}\nabla_{\sigma}\xi)R^{\mu\rho\nu\sigma}.
\end{eqnarray}
In the  FRW universe case, these equations and the equation of motion for the scalar become \cite{N3}
\begin{equation}
0=-\frac{3}{k^2}H^2+\frac{1}{2}\dot{\phi}^2+V(\phi)+24H^3\dot{\xi}+\rho_{m},
\label{4.3}
\end{equation}
\begin{equation}
0=\frac{1}{k^2}(2\dot{H}+3H^2)+\frac{1}{2}\dot{\phi}^2-V(\phi)-8H^2\ddot{\xi}-16H(\dot{H}+H^2)\dot{\xi}+p_{m},
\end{equation}
\begin{equation}
0=\ddot{\phi}+3H\dot{\phi}+V_{\phi}^{'}+\xi_{\phi}^{'}G.
\label{4.5}
\end{equation}
Note that, as a consequence of the Bianchi identity, Eq.~(\ref{4.5}) is satisfied automatically.  From the Eqs.~(\ref{4.3}-\ref{4.5}), we get \cite{N3}
\begin{equation}
0=\frac{2}{k^2}\dot{H}+\dot{\phi}^2-8a(\frac{H^2}{a}\dot{\xi})_t.
\end{equation}
For the variables $\phi$ and $\xi$  Eqs.~(\ref{4.3}-\ref{4.5}) constitute a system of second order differential equations. It is useful to reduce this system to a first order set of differential equations. This can be immediately achieved by introducing new variables, as
\begin{equation}
u=\dot{\phi}^2, \quad v=\dot{\xi},
\end{equation}
Then, we obtain  the following first order system 
\begin{equation}
0=-\frac{3}{k^2}H^2+\frac{1}{2}u+V+24H^3v,
\label{4.8}
\end{equation}
\begin{equation}
0=\frac{1}{k^2}(2\dot{H}+3H^2)+\frac{1}{2}u-V-8H^2\dot{v}-16H(\dot{H}+H^2)v,
\end{equation}
\begin{equation}
0=\dot{u}+6Hu+2\dot{V}+2Gv.
\label{4.10}
\end{equation}
From here it is easy to explore the stability and number of attractors of the physical model. To do so we rewrite the system (\ref{4.8}-\ref{4.10})  as
\begin{equation}
\dot{H}=\frac{1}{24}GH^{-2}-H^2,
\label{4.11}
\end{equation}
\begin{equation}
\dot{u}=-(6Hu+2\dot{V}+2Gv),
\end{equation}\begin{equation}
\dot{v}=\frac{1}{8H^2}[\frac{2}{k^2}(\frac{1}{24}GH^{-2}-H^2)+u-(16H\dot{H}-8H^3)v].
\label{4.13}
\end{equation}
Let $q=u+2V$. Then this can be put as
\begin{equation}
0=-\frac{3}{k^2}H^2+\frac{1}{2}q+24H^3v,
\end{equation}
\begin{equation}
0=\frac{1}{k^2}(2\dot{H}+3H^2)+\frac{1}{2}q-2V-8H^2\dot{v}-16H(\dot{H}+H^2)v,
\end{equation}
\begin{equation}
0=\dot{q}+6Hq-12HV+2Gv,
\end{equation}
and for (\ref{4.11}-\ref{4.13})
\begin{equation}
\dot{H}=\frac{1}{24}GH^{-2}-H^2,
\end{equation}
\begin{equation}
\dot{q}=12HV-6Hq-2Gv,
\end{equation}\begin{equation}
\dot{v}=\frac{1}{8H^2}[\frac{2}{k^2}(\frac{1}{24}GH^{-2}-H^2)+q-2V-(16H\dot{H}-8H^3)v].
\end{equation}
Let us now rewrite Eqs.~(\ref{4.3}-\ref{4.5}) in terms of the number of e-foldings $N$. We have
\begin{equation}
0=-\frac{3}{k^2}H^2+\frac{1}{2}H^2\phi_N^2+V(\phi)+24H^4\xi_{1N},
\label{4.20}
\end{equation}
\begin{eqnarray}
0=\frac{1}{k^2}(2HH_N+3H^2)+\frac{1}{2}H^2\phi_N^2-V(\phi)\nonumber\\
-8H^4\xi_{1NN}-8H(2HH_N+3H^2)H\xi_{1N},
\end{eqnarray}
\begin{equation}
0=H^2\phi_N\phi_{NN}+(H_{N}+3H^2)\phi_N^2+V_N +\xi_{1N}G.
\label{4.22}
\end{equation}
If we introduce new variables
 \begin{equation}
h=\phi_N^2=H^{-2}u, \quad p=\xi_{1N}=H^{-1}v,
\end{equation}
 the system (\ref{4.20}-\ref{4.22}) takes the form
\begin{equation}
0=-\frac{3}{k^2}H^2+\frac{1}{2}H^2h+V(\phi)+24H^4p,
\end{equation}
\begin{equation}
0=\frac{1}{k^2}(2HH_N+3H^2)+\frac{1}{2}H^2h-V-8H^4p_{N}-8H(2HH_N+3H^2)Hp,
\end{equation}
\begin{equation}
0=H^2h_{N}+2(H_{N}+3H^2)h+2V_N +2Gp.
\end{equation}
To summarize, the original system of equations has been written in several ways that can be used to construct solutions. Needless to say, all the corresponding solutions are equally valid in every form of the system. Here we will find solutions for  the one given by Eqs.~(\ref{4.8}-\ref{4.10}). To do tat, we rewrite them in terms of the e-folding variable $N$: 
\begin{equation}
0=-\frac{3}{k^2}H^2+\frac{1}{2}u+V+24H^3v,
\end{equation}
\begin{equation}
0=\frac{1}{k^2}(2HH_{N}+3H^2)+\frac{1}{2}u-V-8H^3v_{N}-16H^2(H_{N}+H)v,
\end{equation}
\begin{equation}
0=Hu_{N}+6Hu+2HV_{N}+2Gv.
\end{equation}
This has the following solution
\begin{equation}
u=8H^4W_{NN}+8H^3(3H_{N}-H)W_{N}-\frac{2}{k^2}HH_{N},
\label{4.30}
\end{equation}
\begin{equation}
v=HW_{N},
\end{equation}
\begin{equation}
V=-4H^4W_{NN}-(12H^2H_{N}+20H^3)HW_{N}+\frac{3}{k^2}H^2+\frac{1}{k^2}HH_{N},
\label{4.32}
\end{equation}
where $W=W(N)$ is some function of $N$. At the same time, $\phi, \xi_1$ are given by
\begin{equation}
\phi=\int \sqrt{8H^2W_{NN}+8H(3H_{N}-H)W_{N}-\frac{2}{k^2}(\ln{H})_{N}}dN,
\end{equation}
\begin{equation}
\xi=W.
\end{equation}
 
 Let us now consider several examples.
\begin{enumerate}
 \item As the first one consider the case when  $W=0$, which corresponds to the Einstein-scalar gravity. In this case, we have
 \begin{equation}
u=-\frac{2}{k^2}HH_{N},\quad v=0,\quad V=\frac{3}{k^2}H^2+\frac{1}{k^2}HH_{N}.
\end{equation}
\item A second  example is  the model
\begin{equation}
W=\nu,
\end{equation}
where $ \nu=const.$ Here the solution of the system (4.30)-(4.32) is 
\begin{equation}
u=-\frac{2}{k^2}HH_{N}, \quad v=0, \quad V=\frac{3}{k^2}H^2+\frac{1}{k^2}HH_{N}.
\end{equation}
It corresponds to the so-called Einstein-Gauss-Bonnet gravity.
\item The third example will be
\begin{equation}
W=\mu N+\nu
\end{equation}
where $\mu, \nu$ are some constants. Then, the solution of the system (4.30)-(4.32) is given by
\begin{equation}
u=8\mu H^3(3H_{N}-H)-\frac{2}{k^2}HH_{N},
\end{equation}
\begin{equation}
v=\mu H (=\mu \dot{N}),
\end{equation}
\begin{equation}
V=-\mu(12H^2H_{N}+20H^3)H+\frac{3}{k^2}H^2+\frac{1}{k^2}HH_{N}.
\end{equation}

The next step is finding the explicit forms of $\phi(t), H(t), \xi_{1}(t), V(t)$. To this end we consider the case when $\phi$ has a kink form, that is, it obeys the (0+1)-dimensional Sine-Gordon equation
\begin{equation}
\ddot{\phi}=\gamma^2\sin{\phi},
\end{equation}
where $\gamma$=const and
\begin{equation}
\phi=4arctg [e^{-\gamma(t-t_0)}], \quad
u=2\gamma^2(1-\cos{\phi})=\frac{4\gamma^2}{1+e^{2\gamma(t-t_0)}}.
\end{equation}
In this case $W$ obeys the equation
\begin{equation}
8H^4W_{NN}+8H^3(3H_{N}-H)W_{N}-\frac{2}{k^2}HH_{N}=2\gamma^2(1-\cos{\phi}).
\end{equation}
We will solve this equation for the case (4.42) and $H=\alpha+\beta e^{mN}$. There, $y$ satisfies
\begin{equation}
\epsilon y^2+\delta y +\sigma=(12\mu\eta-8\mu)y^2+(12\mu\eta-16\alpha\mu-\frac{\eta}{k^2})y-(u+8\mu\alpha^2)=0,
\end{equation}
which has the solution
\begin{equation}
y(t)=y(\phi)=y_{i}=\frac{\delta\pm\sqrt{\delta^2-4\epsilon\sigma}}{2\epsilon}.
\end{equation}
Hence, we have
\begin{equation}
N=N_{i}(t)=N_{i}(\phi)=\frac{1}{\eta}\ln{\frac{y_{i}}{\beta}}
\end{equation}
and
\begin{equation}
H=H(t)=H(\phi)=\sqrt{\alpha+y_{i}}.
\end{equation}
\end{enumerate}
We have thus seen that, from the action (\ref{4.1}), any of the usual cosmologies can be achieved. The system of equations (\ref{4.30}-\ref{4.32}) provides a quite simple setup to reproduce any kind of cosmological solution, as it has been clearly illustrated with the three examples above, where for a given Hubble parameter the scalar Gauss-Bonnet theory is constructed.

\section{Conclusions}

In the present paper, several types of DE cosmologies in modified GB gravity---which can be viewed as being inspired by string considerations, Ref.\cite{N3}---have been investigated. We have studied different kinds of theories, in all of which the GB invariant plays an important role in the corresponding equations. First, we have shown that GR plus some function of the GB term provides a very powerful theory, where no sort of dark energy is actually needed to reproduce the standard $\Lambda$CDM cosmology. 

More general theories have been considered too, as in Sect.~III, where  an action depending on a function of the Ricci scalar and GB invariant has been studied. In this case the dark energy behavior can also be reproduced, while its extra degrees of freedom provide us with a powerful tool to constraint the theory, in order to avoid violation of the local gravity tests. 

In the same way, the scalar GB theory studied in the last section adequately reproduces any kind of cosmological solution. As a follow up, we have shown, with the help of several particular examples corresponding to explicit choices of the functions $f(G)$, $F(R,G)$ or the scalar field that, in principle, any cosmic evolution can be obtained from these models, what includes the unification of early-time inflation with the late-time acceleration coming from astronomical observations. To finish, it has been indicated that DE cosmologies in a more general (and complicated) $F(R,G)$ framework---generically requiring numerical analysis---can be reconstructed in a similar fashion, too.

\ack
DSG acknowledges a grant from MICINN (Spain), project FIS2006-02842.
Part of EE's research was performed while on leave at Department of Physics and
Astronomy, Dartmouth College, 6127 Wilder Laboratory, Hanover, NH 03755,
USA. This work was also supported by MEC (Spain) grant PR2009-0314, and by AGAUR (Generalitat de Catalunya), contract 2009SGR-994.

\end{document}